\documentclass[11pt]{amsart}

\usepackage[latin2]{inputenc}
\usepackage{amsmath}
\usepackage{amsthm}
\usepackage{amsfonts}

\begin{document}

\newtheorem{theorem}{Theorem}
\newtheorem{lemma}[theorem]{Lemma}
\newtheorem{claim}[theorem]{Claim}
\newtheorem{fact}[theorem]{Fact}
\newtheorem{problem}[theorem]{Problem}

\def\chrom{\textsc{chromatic number}}
\def\kcol{$k$\textsc{-colorability}}
\def\klistcol{$k$\textsc{-list-coloring}}

\def\sgn{\mathop{\rm sgn}}
\def\cval{\mathop{\rm cval}}

\title[Polynomial-time algorithm for vertex $k$-colorability of $P_5$--free graphs]{Polynomial-time algorithm for vertex \\ $k$-colorability  of $P_5$--free graphs}

\author[Marcin Kaminski and Vadim Lozin]{Marcin~Kami\'nski$^*$ and Vadim Lozin$^\S$\\\\}

\thanks{\hspace*{-.5cm} $^*$RUTCOR, Rutgers Univeristy, 640 Bartholomew Road, Piscataway, NJ 08854,
USA, \\email: mkaminski@rutcor.rutgers.edu \\ $^\S$~RUTCOR, Rutgers
Univeristy, 640 Bartholomew Road, Piscataway, NJ 08854, USA,\\
email: lozin@rutcor.rutgers.edu}



\begin{abstract}
We give the first polynomial-time algorithm for coloring vertices of
$P_5$-free graphs with $k$ colors. This settles an open problem and
generalizes several previously known results.
\end{abstract}

\maketitle

\section{Introduction}
A \emph{$k$-coloring} of a graph is an assignment of numbers from
set $[k] := \{1, \ldots, k\}$ (called \emph{colors}) to the vertices
of a graph in such a way that the endpoints of each edge receive
different colors.

It is well known that deciding $k$-colorability is an NP-complete
problem for any $k\ge 3$. Moreover, it remains difficult under
substantial restrictions, for instance, for line graphs
\cite{Degree4} (which is equivalent to edge colorability), graphs of
low degree \cite{Nbrhds} or triangle-free graphs \cite{MP}. On the
other hand, the problem can be solved in polynomial time for perfect
graphs \cite{GLS84}, locally connected graphs \cite{Loco},  and for
some classes defined by forbidden induced subgraphs
\cite{Ran04,RS04,RST02}.

Given a class of graphs without induced subgraphs isomorphic to the
induced path on $t$ vertices (we denote such a path by $P_t$ and
call such graphs $P_t$-free), we want to investigate whether the
\kcol~problem can be solved in this class in polynomial time or can
be proved to be $NP$-complete. A number of results has been obtained
in this area for different combinations of parameters $k$ and $t$.

Sgall and Woeginger showed in \cite{SW01} that
$5$\textsc{-colorability} is NP-complete for $P_8$-free graphs and
$4$\textsc{-colorability} is NP-complete for $P_{12}$-free graphs.
The last result was improved in \cite{LRS06}, where the authors
claim that modifying the reduction from \cite{SW01}
$4$\textsc{-colorability} can be shown to be NP-complete for
$P_9$-free graphs.

The problem can be solved in polynomial time for $P_4$-free graphs
as they constitute a subclass of perfect graphs. Two more
polynomial-time results are deciding $3$-{\sc colorability} of
$P_5$-free \cite{RST02,SW01} and $P_6$-free graphs \cite{RS04A}.

Table \ref{complexity_before} summarizes known results for the
\kcol~problem in the class of $P_t$-free graphs. (A similar table
appears for the first time in \cite{SW01} and is being updated and
redrawn ever since in all publications contributing to the area.)
When looking at the previous results, two possible research
directions seem promising and they were listed as open problems in
\cite{RS04}. We restate them here.

\begin{table}[ht]
\begin{tabular}{|c||c|c|c|c|c|c|c|c|}
  \hline
   & $t = 4$ & $t = 5$ & $t = 6$ & $t = 7$ & $t = 8$ & $t = 9$ & $t = 10$  &  \,\, \ldots \,\, \\
   \hline \hline
   $k = 3$ & P & P & P & ? & ? & ? & ? & ? \\  \hline
   $k = 4$ & P & ? & ? & ? & ? & NPc & NPc & NPc \\  \hline
   $k = 5$ & P & ? & ? & ? & NPc & NPc & NPc & NPc \\  \hline
   $k = 6$ & P & ? & ? & ? & NPc & NPc & NPc & NPc \\  \hline
   $k = 7$ & P & ? & ? & ? & NPc & NPc & NPc & NPc \\  \hline
   \ldots & P & ? & ? & ? & NPc & NPc & NPc & NPc \\  \hline
\end{tabular}
\vspace*{.2cm}
  \caption{Complexity results for $k$-colorability of $P_t$-free
  graphs.} \label{complexity_before}
\end{table}

\begin{problem}
Is there an polynomial-time algorithm for the
$4$\textsc{-colorability} problem in the class of $P_5$-free graphs?
\end{problem}

\begin{problem}
Is there an polynomial-time algorithm for the
$3$\textsc{-colorability} problem in the class of $P_7$-free graphs?
\end{problem}

As far as we know, there has been no progress on the second problem,
while for the first some partial results were obtained. The authors
of \cite{LRS06} showed that the $4$\textsc{-colorability} problem
can be solved in polynomial time in the class of $(P_5, C_5)$-free
graphs. Another result was obtained in \cite{HSW06}, where the
authors present a polynomial-time algorithm to solve the
$4$\textsc{-colorability} problem in $P_5$-free graphs containing a
dominating clique on four vertices.

\medskip
In this paper, we give a complete solution to the \kcol~problem in
$P_5$-free graphs for an arbitrary value of $k$. In fact, our
algorithm solves a more general version of the problem, known as
{\sc list coloring}. We formally define this problem and provide
other necessary background information in the next section.

\section{Preliminaries}

The algorithmic problem we study in this paper is \klistcol. An
instance of the problem consists of a graph and a list of colors
available for each vertex. More formally, an instance
$G=(V,E,\mathcal{L})$ of the \klistcol~ problem is a graph with
vertex set $V$, edge set $E$, and a function $\mathcal{L} : V
\rightarrow 2^{[k]}$. An instance $G$ is $k$-colorable if there
exists a $k$-coloring of the vertices of $G$ such that each vertex
$v$ is assigned a color from $\mathcal{L}(v)$. A $k$-coloring is
called {\it chromatic} if the graph is not $(k-1)$-colorable.

For a set $W \subseteq V$, we write $\mathcal{L}(W) = \bigcup_{w \in
W} \mathcal{L}(w)$. We say that $\mathcal{L}(v)$ (or
$\mathcal{L}(W)$) is the \emph{palette} of $v$ (or $W$). When we
want to emphasize the underlying instance $H$, we write
$\mathcal{L}_H$; if the subscript is omitted, we always refer to $G$.

If a vertex is assigned a color, we can exclude that color from the
palettes of all its neighbors. We say that an instance $G = (V, E,
\mathcal{L})$ is in simplified form if there are no adjacent
vertices $v,w \in V$ such that $|\mathcal{L}(v)| = 1$ and
$\mathcal{L}(v) \subseteq \mathcal{L}(w)$. In this paper we assume
that every instance is in simplified form and that instance
simplification in the algorithm is done implicity. (It can be easily
performed in time linear in the number of edges.)

While looking for a coloring in a graph, two adjacent vertices with
disjoint palettes can be as well thought of as non-adjacent.
Essential are only those pairs of adjacent vertices whose palettes
are not disjoint. This observation motivates the following
definition. For~a~vertex $v \in V$, we define the set of its
\emph{essential neighbors} $\mathcal{E}(v) = \{ w \in N(v) \, : \,
\mathcal{L}(v) \cap \mathcal{L}(w) \neq \emptyset \}$. The remaining
neighbors $N(v) - \mathcal{E}(v)$ are called \emph{non-essential}.
Similarly, for a set $W \subseteq V$, we define the set of its
essential neighbors $\mathcal{E}(W) = \bigcup_{v \in W}
\mathcal{E}(v)$. Note that the relation of being an essential
(non-essential) neighbor is symmetric. Also, assigning a color to a
vertex does not change possible color choices for its non-essential
neighbors.

\medskip

Our solution is based on an interesting structural property of
$P_5$-free graphs that has been described by Basc\'o and Tuza in
\cite{BT90}. (The properties of graphs without long induced paths
have been also studied in other papers: \cite{BT90A, BT93, BT02,
CK90}.) Following their terminology, we say that a graph $H$ is dominating
in $G$ if $G$ contains a dominating set that induces a graph isomorphic to $H$. In
particular, a dominating clique in $G$ is a dominating set
which induces a complete graph. Similarly, a dominating $P_3$ is a
dominating set which induces a path on 3 vertices.

\begin{theorem}
[\cite{BT90}] In every $P_5$-free connected graph there
is a dominating clique or a dominating $P_3$.
\label{dominating_structure_thm}
\end{theorem}

We will refer to a dominating set that induces a complete graph or
$P_3$ as a \emph{dominating structure}. To give an application of
the theorem, let us consider the $3$\textsc{-list-coloring} problem
in the class of $P_5$-free graphs. Notice that once a $3$-coloring
of vertices in a dominating structure $D$ is fixed, then
$|\mathcal{L}(v)| \leq 2$ for all $v \in V$. The question whether
the coloring of $D$ can be extended to the whole graph, can be
modeled as a $2$-\textsc{SAT} instance and solved in polynomial
time. Hence, considering all possible $3$-colorings of $D$ and
checking extendability of each, we can obtain a polynomial algorithm
for the $3$\textsc{-list-coloring} problem in the class of
$P_5$-free graphs. (See \cite{RST02} for more details.)

Taking into account the special role of dominating sets in
$P_5$-free graphs, we extend each instance of the \klistcol~problem
by adding to it a dominating structure, i.e., throughout the paper
an instance $G=(V,E,\mathcal{L},D)$ of the \klistcol~ problem is a
graph $G$ with vertex set $V$, edge set $E$, function $\mathcal{L} :
V \rightarrow 2^{[k]}$ and a dominating set $D \subseteq V$.

Given a dominating set $D$, we partition all vertices in $V - D$ into disjoint subsets
depending on their neighborhood in the set $D$. For $I
\subseteq D$, $U_I(G) = \{ v \in V \setminus D \, : \, N(v) \cap D =
I \}$. If $G$ is the underlying instance, we just write $U_I$. The
sets $U_I$ for $I \subseteq D$ will be referred to as \emph{bags}.

The nature of our solution is inductive. Designing the algorithm
that solves the \klistcol~problem, we assume there exist
polynomial-time algorithms for the same problem with smaller values
of $k$. The problem can be easily solved for $k = 1,2$ and, with a
bit more effort, for $k=3$. So, below we assume that $k$ is at least
$4$. Notice that our inductive assumption allows us to find a
chromatic coloring of a $(k-1)$-colorable $P_5$-free graph in
polynomial time.

The main idea of our $k$-coloring algorithm is to use the structural
property of $P_5$-free graphs to create a set $\mathcal{G}$ of
simpler instances. We will say that an instance $G$ is compatible
with a set $\mathcal{G}$ if $G$ is $k$-colorable if and only if at
least one of the instances in $\mathcal{G}$ is $k$-colorable. Notice
that if $G$ is compatible with $\mathcal{G}$ and some $H \in
\mathcal{G}$ is compatible with $\mathcal{H}$, then $G$ is
compatible with $(\mathcal{G} \setminus H) \cup \mathcal{H}$. For
clarity, we divided the description of our solution into three
parts, each corresponding to one of the following sections.
Throughout the paper, $n$ stands for the number of vertices of $G$,
and `polynomial' means `polynomial in $n$'.

\section{Dominating an independent set}
\label{sec_dis}
Let $G=(V,E,\mathcal{L},D)$ be an instance of the problem, $U_I,
U_J$ two different bags, and $S, T$ two independent sets belonging
to $U_I$ and $U_J$, respectively. We denote by $S'$ the set of
essential neighbors of $T$ in $S$, and similarly, by $T'$ the set of
essential neighbors of $S$ in $T$. Observe that $S'$ is empty if and
only if $T'$ is empty. Moreover, every vertex of $S'$ has a neighbor
in $T'$ and vice versa.

\begin{lemma}\label{dominating_vertex}
If $S' \neq \emptyset$, there exists a vertex in $S'$ that is
adjacent to all vertices in $T'$.
\end{lemma}

\noindent \emph{Proof}. Let $s_1$ be a vertex of $S'$ with a maximal
neighborhood in $T'$. Assume there exists a vertex $t_2 \in T'$ that
is not adjacent to $s_1$. Then, there must exist a vertex $s_2 \in
S'$ (different than $s_1$) adjacent to $t_2$. By the choice of
$s_1$, the vertex $s_2$ must have a non-neighbor $t_1 \in T' \cap
N(s_1)$. Since $I \neq J$, there exists a vertex $v \in (I \setminus
J) \cup (J \setminus I)$, but then $G[v, s_1, s_2, t_1, t_2]$ is an
induced $P_5$; a contradiction. \qed \vspace*{.2cm}

Notice that from Lemma~\ref{dominating_vertex} it follows that the
vertices of $S'$ can be linearly ordered with respect to the
neighborhood containment.

Let $v \in S'$ be a vertex that dominates $T'$ and let us look at
the palette of $v$. We can divide it into two parts -- the colors
that belong to the palette of $T'$ and the remaining ones. Notice
that assigning to $v$ one of the colors from the palette of $T'$
decreases the size of the palette of $T'$ in the resulting instance.
On the other hand, truncating the palette of $v$ so that it contains
only the colors not belonging to the palette of $T'$ decreases the
size of $S'$ in the resulting instance. The following procedure
makes use of this observation.

\medskip
\medskip
\fbox{ \begin{minipage}[b]{14cm}
\vspace*{.1cm}
{\sc Procedure $\Pi_{S, T}$}
\vspace*{.4cm}

\noindent {\sc Input}: Instance $G = (V, E, \mathcal{L}, D)$.
\vspace*{.1cm}

\noindent {\sc Output}: Set $\mathcal{G}$ of instances. \vspace*{.4cm}

\noindent {\sc Step 1}. Let $\mathcal{G} = \emptyset$. If $S' =
\emptyset$, then {\sc Return $\{ G \}$}. \vspace*{.15cm}

\noindent {\sc Step 2}. Find a vertex $v \in S'$ that dominates
$T'$. For every $d \in \mathcal{L}(v) \cap \mathcal{L}(T')$, {\sc
Add to $\mathcal{G}$} an instance $G' = (V, E, \mathcal{L}_{G'}, D)$
such that $\mathcal{L}_{G'}(v) = d$ and $\mathcal{L}_{G'}(w) =
\mathcal{L}(w)$ for all vertices $w \in V \setminus \{v\}$.
\vspace*{.15cm}

\noindent {\sc Step 3}. If $ \mathcal{L}(v) \setminus
\mathcal{L}(T') \neq \emptyset$, create an instance $G' = (V, E,
\mathcal{L}_{G'}, D)$ such that $\mathcal{L}_{G'}(v) =
\mathcal{L}(v) \setminus \mathcal{L}(T')$ and $\mathcal{L}_{G'}(w) =
\mathcal{L}(w)$ for all vertices $w \in V \setminus \{v\}$. {\sc Add
to $\mathcal{G}$} the instances returned by $\Pi(G')$.
\vspace*{.15cm}

\noindent {\sc Step 4}. {\sc Return $\mathcal{G}$}. \vspace*{.15cm}

\end{minipage}}

\begin{claim}
Let $G$ be the input instance and $\mathcal{G}$ the output set of instances of
{\sc Procedure $\Pi_{S, T}$}. Then
\begin{itemize}
\item[(*)] $\mathcal{G}$ compatible with $G$ and
\item[(**)] for each $G_t \in \mathcal{G}$, $S'_{G_t} = \emptyset$ or $|
\mathcal{L}_{G_t}(T'_{G_t})| < | \mathcal{L}(T') |$.
\end{itemize}
Moreover, {\sc Procedure $\Pi_{S, T}$} runs in polynomial time.
\end{claim}

\noindent \emph{Proof}. To prove (*) and (**), we will proceed by induction on $|S'|$.

If $|S'| = 0$, then $T' = \emptyset$ and $\mathcal{G}$ consists only
of $G$ ({\sc Step 1}). Clearly, $G$ is compatible with $\{G\}$ and
(**) is also satisfied.

Suppose that $|S'| = i$ and for all instances $H$ with $|S'_H| < i$
the output of the procedure satisfies both conditions (*) and (**).
First, let us notice that if one of the instances in $\mathcal{G}$
is $k$-colorable, then so is $G$ because for each instance $G_t \in
\mathcal{G}$ and each vertex $v \in V$, $\mathcal{L}_{G_t}(v)
\subseteq \mathcal{L}(v)$.

Now suppose that $G$ is $k$-colorable. Since in any $k$-coloring of
$G$, $v$ receives a color from $\mathcal{L}(v)$, then either one of
instances created in {\sc Step 2} or $G'$ is $k$-colorable. If none
of the instances created in {\sc Step~2} is $k$-colorable, then $G'$
must be $k$-colorable and -- by the induction hypothesis -- at least
one of the instances created in {\sc Step 3} is $k$-colorable.
Hence, (*) is satisfied.

It is easy to see that all instances $G_t$ created in {\sc Step 2}
have $|\mathcal{L}_{G_t}(T'_{G_t})| < | \mathcal{L}(T')|$. The set
of instances created in {\sc Step 3} comes from a call of the
procedure for $G'$ and for these instances the condition is
satisfied by the induction hypothesis, since $|S'_{G'}| < |S'|$.
Hence, (**) is satisfied.

Now let us show the running time of the procedure. Clearly,
identifying sets $S', T'$, finding a dominating vertex $v$ ({\sc
Step~2}) and creating new instances can be done in polynomial time.
Notice that the recursive call in {\sc Step~3} is done for an
instance with a smaller essential part of $S$ so the depth of the
recursion is at most $n$. Hence, the running time follows.\qed
\vspace*{.2cm}

Now we are going to use {\sc Procedure $\Pi_{S, T}$} to design an algorithm that
given $G$ creates a set of instances $\mathcal{G}$ compatible with
$G$. We also want $\mathcal{G}$ to have a polynomial size and we
require that in each instance $G_t \in \mathcal{G}$,
$S$ has no essential neighbors in $T$.

\begin{lemma}
There exists a polynomial-time {\sc Algorithm} $\Pi'_{S, T}(G)$ such
that given two independent sets $S \subseteq U_I^J$ and $T \subseteq
U_J^I$ generates a set of instances $\mathcal{G}$ compatible with
$G$, such that for each $G_t \in \mathcal{G}$, $S'_{G_t} =
\emptyset$.
\end{lemma}

\noindent \emph{Proof}. Each call of {\sc Procedure $\Pi_{S, T}$}
produces a set compatible with $G$ that has a polynomial number of
members (in fact at most $kn$). All members have either $S'(G_t) =
\emptyset$ or fewer colors in the palette of $T'(G_t)$ than in the
palette of $T'$.

Calling {\sc Procedure $\Pi_{S, T}$} recursively until all instances
have the property $T'(G_t) = \emptyset$ builds a search tree of
bounded depth (at most $k$) and polynomial degree (at most $kn$).
Hence, the number of instances is polynomial and so is the running
time of the algorithm. \qed

\section{Dominating color classes}
\label{sec_dcc}

In this section, as in the previous one, $G=(V,E,\mathcal{L},D)$ is an instance of the problem,
and $U_I, U_J$ are two different bags. We denote by $U_I^J$ the set of essential neighbors of $U_I$ in $U_J$.
{\sc Procedure $\Theta_{I, J}$} presented in this section is parameterized by $I$
and $J$.

\fbox{ \begin{minipage}[h]{13.5cm} \vspace*{.1cm} {\sc PROCEDURE
$\Theta_{I, J}$} \vspace*{.4cm}

\noindent {\sc Input}: Instance $G = (V, E, \mathcal{L}, D)$.
\vspace*{.1cm}

\noindent {\sc Output}: Set $\mathcal{G}$ of instances. \vspace*{.4cm}

\noindent {\sc Step 1}. If $U_I^J = \emptyset$, {\sc Return
$G$}. \vspace*{.15cm}

\noindent {\sc Step 2}. Find a chromatic coloring of $G[U_I^J]$ and
let $A$ be one of the color classes (non-empty). Color $G[U_J^I]$
with $k-1$ colors and let $B_1, \ldots, B_{k-1}$ be the color
classes of that coloring. If $G[U_I^J]$ or $G[U_J^I]$ are not
$k$-colorable, then {\sc Return $\{\emptyset\}$}. Otherwise, let
$\mathcal{G} := \{ G \}$ and $\mathcal{H} := \emptyset$.
\vspace*{.15cm}

\noindent {\sc Step 3}. {\sc For each} $i = 1, \ldots, k$ {\sc Do}
\vspace*{.15cm}

\noindent {\sc Step 4}. \hspace{.5cm} {\sc If} $B_i \neq \emptyset$
{\sc Then} \vspace*{.15cm}

\noindent {\sc Step 5}. \hspace*{1cm} {\sc For each} $G_t \in
\mathcal{G}$, \\ \hspace*{2.5cm} {\sc Add to $\mathcal{H}$} the
instances returned by $\Pi'_{A, B_i}(G_t)$. \vspace*{.15cm}

\noindent {\sc Step 6}. \hspace*{1cm} $\mathcal{G} := \mathcal{H}$,
$\mathcal{H} := \emptyset$. \vspace*{.15cm}

\noindent {\sc Step 7}. {\sc End For} \vspace*{.15cm}

\noindent {\sc Step 8}. {\sc Return $\mathcal{G}$}. \vspace*{.15cm}

\end{minipage}}

\begin{claim}
Let $G$ be the input instance and $\mathcal{G}$ the output set of instances of
{\sc Procedure $\Theta_{I, J}$}. Then $G$ is
compatible with $\mathcal{G}$ and for each $G_t \in \mathcal{G}$,
either $U_I^J(G_t) = \emptyset$ or the chromatic number of $U_I^J(G_t)$ is strictly smaller than that of $U_I^J(G)$.
Moreover, {\sc Procedure $\Theta_{I, J}$} runs in polynomial time.
\end{claim}

\noindent \emph{Proof}.
First let us notice that the set $\mathcal{H}$ is obtained from the
set $\mathcal{G}$ by replacing instances $G_t \in \mathcal{G}$ with
a set of instances compatible with $G_t$. Hence, after {\sc Step 5},
$G$ is compatible with the set $\mathcal{H}$ if and only if $G$ is
compatible with the set $\mathcal{G}$. Since $G$ is compatible with
$\mathcal{G}$ before the loop ({\sc Step 2}), it is also compatible
after {\sc Step 8}, and therefore $G$ is compatible with the output
set $\mathcal{G}$.

Notice that after $i$-th iteration of the loop ({\sc Steps 3 -- 7}),
for all $G_t \in \mathcal{G}$ there are no vertices in $A$ that have
an essential neighbor in $B_i$. Therefore at {\sc Step 8}, for all
$G_t \in \mathcal{G}$, no vertex from $A$ has an essential neighbor
in $U_J^I$ and clearly either $U_I^J(G_t) = \emptyset$ or
$\chi(U_I^J(G_t)) < \chi(U_I^J)$.

Each call of {\sc Procedure} $\Pi'_{A, B_i}$ in {\sc Step 5}
produces a polynomial number of instances with a smaller chromatic
number. For each such an instance the procedure $\Pi'$ is called
recursively and since the depth of the recursion is bounded by $k$,
the running time of the algorithm is polynomial. \qed \vspace*{.2cm}

Now we use the procedure to design an algorithm that given $G$
creates a set of instances $\mathcal{G}$ compatible with $G$ such
that for each $G_t \in \mathcal{G}$, $U_I^J(G_t)$ is empty.

\begin{lemma} \label{dominate_U_I}
There exists a polynomial-time algorithm $\Theta'_{I, J}$ that given
two different sets $I, J \subset D$ generates a set of instances
$\mathcal{G}$ compatible with $G$ such that for each $G_t \in
\mathcal{G}$, $U_I^J(G_t) = \emptyset$.
\end{lemma}

\noindent \emph{Proof}. Calling {\sc Procedure $\Theta_{I, J}$}
recursively until instances have the property $U_I^J(G_t) =
\emptyset$ builds a search tree of bounded depth (at most $k$),
since at each step the chromatic number of $U_I^J(G_t)$
decreases. The degree of each node in this tree is bounded by a polynomial. Hence, the running time of the
algorithm is polynomial.
\qed

\section{Main algorithm} \label{sec_main}

In this section we combine techniques described above to construct
an algorithm that solves the \klistcol~problem in the class of
$P_5$-free graphs. Let us notice that we can assume that the input
graph is connected, as if it is not, the \klistcol~problem can be
solved on its connected components separately.

We divide the presentation of the main algorithm into three steps.
First, we make a simple observation about an instance whose all bags
are separated. A bag is called \emph{separated} if all essential
neighbors of its vertices belong to the bag itself.

\begin{lemma}
Let $G = (V, E, \mathcal{L}, D)$ be an instance of the
\klistcol~problem such that for each $I \subset D$,
the bag $U_I$ is separated and for each
$v \in D$, $|\mathcal{L}(v)| = 1$. The \klistcol~problem can be solved on $G$ in polynomial
time. \label{separated_bags}
\end{lemma}

\noindent \emph{Proof}. Since vertices of $U_I$ have no essential
neighbors outside $U_I$ and vertices of the dominating structure
have been already colored, graphs $G[U_I]$ can be colored separately
for each $I \subset D$ and solutions can be glued together. Notice
that each bag $U_I$ together with the set $I$ dominating it is in fact an instance of the
$(k-1)$\textsc{-list-coloring} problem. By the inductive assumption
this can be solved in polynomial time. \qed \vspace*{.2cm}

Second, we show that there is a polynomial-time procedure that given an
instance $G$ with a dominating set of bounded size creates a set of
instances $\mathcal{G}$ compatible with $G$ such that in each instance all the bags are separated.

\medskip
\medskip
\fbox{ \begin{minipage}[t]{13.5cm} \vspace*{.1cm} {\sc ALGORITHM
$\Lambda$} \vspace*{.4cm}

\noindent {\sc Input}: Instance $G = (V, E, \mathcal{L}, D)$. \vspace*{.1cm}

\noindent {\sc Output}: Set $\mathcal{G}$ of instances.
\vspace*{.4cm}

\noindent {\sc Step 0}. Let $\mathcal{G} := \{ G \}$ and $\mathcal{H} :=
\emptyset$. \vspace*{.15cm}

\noindent {\sc Step 1}. {\sc For each} $k$-coloring of $D$ {\sc Do}
\vspace*{.15cm}

\noindent {\sc Step 2}. \hspace{.5cm} {\sc For each} $I, J \subset
D$, $I \neq J$ {\sc Do} \vspace*{.15cm}

\noindent {\sc Step 3}. \hspace{1cm} {\sc If} $U_I^J \neq \emptyset$
{\sc Then} \vspace*{.15cm}

\noindent {\sc Step 4}. \hspace*{1.5cm} {\sc For each} $G_t \in
\mathcal{G}$, \\ \hspace*{3cm} {\sc Add to $\mathcal{H}$} the
instances returned by $\Theta'_{I, J}(G_t)$; \vspace*{.15cm}

\noindent {\sc Step 5}. \hspace*{1cm} $\mathcal{G} := \mathcal{H}$,
$\mathcal{H} := \emptyset$; \vspace*{.15cm}

\noindent {\sc Step 6}. \hspace*{.5cm}  {\sc End For}
\vspace*{.15cm}

\noindent {\sc Step 7}. {\sc End For} \vspace*{.15cm}

\noindent {\sc Step 8}. {\sc Return $\mathcal{G}$} \vspace*{.15cm}

\end{minipage}}

\begin{lemma}
For any input $G$, the set $\mathcal{G}$ of instances returned by {\sc Algorithm $\Lambda$} is compatible with $G$, and in each instance  all the bags are separated. Moreover, if $|D| \leq k$, then the running time of {\sc Algorithm $\Lambda$} is polynomial.
\end{lemma}

\emph{Proof}. First let us notice that since the size of $D$ is
bounded so is the number of $k$-colorings of $D$ ({\sc Step 1}) and
the number of pairs of subsets $I, J \subset D$ ({\sc Step 2}).
Hence, {\sc Step 3} will be performed at most a constant number of
times and {\sc Step 4} takes a polynomial time, so the polynomial
running time of the whole algorithm follows.

From Lemma \ref{dominate_U_I} it is clear that after {\sc Step 4},
$G$ is compatible with $\mathcal{H}$ if and only if $G$ is
compatible with $G$. After {\sc Step 5}, all graphs $G_t$ in
$\mathcal{G}$ have $U_I^J(G_t) = 0$ for all pairs $I, J$ that have
been considered so far. Clearly, at {\sc Step 8} all instances in
$\mathcal{G}$ have $U_I^J(G_t) = 0$ for all pairs $I, J \subset D$,
$I \neq J$ and, hence, $U_I(G_t)$ is separated for each $I \subset D$. \qed \vspace*{.2cm}

Now we are ready to state our main result.

\begin{theorem}
There exists a polynomial-time algorithm for the \klistcol~problem.
\end{theorem}

\noindent \emph{Proof}. A $k$-colorable graph cannot contain a
clique on $k+1$ vertices as its subgraph. We assume that the input
instance $G$ does not contain such a subgraph. (This can be done in
polynomial time, and if $G$ contains a clique on $k+1$ vertices,
then the instance is not $k$-colorable.)

According to Theorem \ref{dominating_structure_thm}, a connected
$P_5$-free graph contains either a dominating clique or a dominating
$P_3$. Since the size of any clique in $G$ is at most $k$, the size
of the dominating structure is also bounded by $k$ and a dominating
set $D$ can be found in polynomial time.

{\sc Algorithm $\Lambda$} is called for such an instance $G$. It
creates a set of instances $\mathcal{G}$ that is compatible with
$G$. Moreover, all bags of every graph in $\mathcal{G}$ are
separated and instances of this type can be handled by Lemma
\ref{separated_bags}. \qed

\begin{table}
\begin{tabular}{|c||c|c|c|c|c|c|c|c|}
  \hline
   & $t = 4$ & $t = 5$ & $t = 6$ & $t = 7$ & $t = 8$ & $t = 9$ & $t = 10$  &  \,\, \ldots \,\, \\
   \hline \hline
   $k = 3$ & P & P & P & ? & ? & ? & ? & ? \\  \hline
   $k = 4$ & P & P & ? & ? & ? & NPc & NPc & NPc \\  \hline
   $k = 5$ & P & P & ? & ? & NPc & NPc & NPc & NPc \\  \hline
   $k = 6$ & P & P & ? & ? & NPc & NPc & NPc & NPc \\  \hline
   $k = 7$ & P & P & ? & ? & NPc & NPc & NPc & NPc \\  \hline
   \ldots & P & P & ? & ? & NPc & NPc & NPc & NPc \\  \hline
\end{tabular}
\vspace*{.2cm}
  \caption{Improved complexity results for $k$-colorability of $P_t$-free
  graphs.} \label{complexity_now}
\end{table}

\section{Conclusion}

In this paper we gave the first polynomial time algorithm solving
the \klistcol~problem in the class of $P_5$-free graphs. The updated
Table \ref{complexity_now} presents the current landscape of
complexity results on colorability in graphs without long induced
paths.

We purposely do not provide a more precise time bound of the running
time as the recursive nature of our solution makes it highly
exponential in $k$. To find a better fixed parameter algorithm for
the $k$-colorability problem of $P_5$-free graphs is an interesting
direction for further research. Another would be to determine what
is the computational complexity of finding the chromatic number of a
$P_5$-free graph.

\end{document}